# Performance Analysis of Wireless Network with Opportunistic Spectrum Sharing via Cognitive Radio Nodes

R. Kaniezhil and C. Chandrasekar

*Abstract*—Cognitive radio (CR) is found to be an emerging key for efficient spectrum utilization. In this paper, spectrum sharing among service providers with the help of cognitive radio has been investigated. The technique of spectrum sharing among service providers to share the licensed spectrum of licensed service providers in a dynamic manner is considered. The performance of the wireless network with opportunistic spectrum sharing techniques is analyzed. Thus, the spectral utilization and efficiency of sensing is increased, the interference is minimized, and the call blockage is reduced.

## 1. Introduction

Today, wireless communication deals with two main problems: spectrum scarcity and deployment delays. These problems are caused by the centralized manner and static in nature of frequency assignment. This scheme cannot adapt to the changing needs of spectrum by users from the military, governmental, and commercial purposes. New technologies should be used effectively to provide the flexibility for the above said problems.

Spectrum is no longer sufficiently available, because it has been assigned to primary users that own the privileges to their assigned spectrum. However, it is not used efficiently most of the time. In order to use the spectrum in an opportunistic manner and increase the spectrum availability, the unlicensed users can be allowed to utilize licensed bands of licensed users, without causing any interference with the assigned service.

The licensed bands of primary users are allocated to the unlicensed users only under the protocol of no interference with the assigned services. This paradigm for wireless communication is known as an opportunistic spectrum access and this is considered to be a feature of cognitive radio (CR). CR is an emerging wireless communication paradigm in which either the network or the wireless node itself intelligently adapts particular transmission or reception parameters by sensing the environment.

Dynamic spectrum access using CR is an emerging research topic. CR techniques provide the capability to use or share the spectrum in an opportunistic manner.

The Federal Communications Commission (FCC) spectrum policy has recommended a paradigm shift in interference assessment from the largely fixed operations. This facilitates real-time interactions between a transmitter and a receiver in an adaptive manner. The recommendation is based on a new metric called the interference temperature, which is intended to quantify and manage the sources of interference in a radio environment. The interference is defined to be the radio frequency (RF) power measured at a receiving antenna per unit bandwidth. The key ideas for this new metric are as follows.

1) The interference temperature at a receiving antenna provides an accurate measure for the acceptable level of RF interference in the frequency band of interest. Any transmission in that band is considered to be "harmful", if it increases the noise floor above the interference threshold.

2) At a given particular frequency band in which the interference temperature is not exceeded, the band could be made available to secondary users. Hence, a secondary device might attempt to coexist with the primary and the presence of secondary devices goes unnoticed.

We have already proposed the spectral efficiency in our previous works[1]–[4]. Here, we mainly concentrate on minimizing the interference and improving quality of service (QoS) based on spectrum sharing using CR nodes, and the overall spectral efficiency is carried out in the present work.

The paper is organized as follows. Sections 2, 3 and 4 define related works, CR, and dynamic spectrum access for its implementation. In Section 5, a spectrum sharing proposal has been discussed. In Section 6, performance metrics has been investigated to improve the system efficiency. In Section 7, the simulation results and implementation issues are discussed. Finally, conclusions are presented in Section 8.

## 2. Related Work

There have been several research efforts on spectrum sharing in CR technologies in order to avoid spectrum scarcity and improve the spectrum utilization. This is considered as the main goal of our research work in a coordinated distributed manner and long-term spectrum assignment strategy. Many research works were proposed under centralized and decentralized manner of spectrum sharing.

In [5], it was discussed that the spectrum utilization mainly hangs on the percentage of spectrum utilized by one service provider that depends upon number of users served and how much spectrum each user application demands in a coordinated manner centralized network. In [6] and [7], the authors proposed a novel spectrum sharing strategy based on the throughput model in a cognitive radio network, where the maximum theoretical throughput is taken into account. In [8], the author addressed the problems of power control and channel assignment jointly. Here, the author devised a near-optimal yet simple algorithm with linear complexity targeting capacity maximization of a CR network while jointly optimizing power and channel allocation among users. In [9], the authors proposed a novel multi-channel connection graph method for achieving a fair spectrum sharing. In [10], spectrum efficiency in multi-hop transmissions was needed by considering the route selection and spectrum management.

The differences between this work and previous works are summarized as follows. First of all, the resource allocation in a cognitive wireless network is quite different from that in traditional wireless networks such as 802.11 based wireless networks due to its special features, like dynamic channel availability, channel heterogeneity, and so on[11]. Second, fairness is considered as a major one of this work. However, the proposed work achieves different goals such as minimizing active users when there is a high traffic, reducing call blockage, and maximizing system efficiency. Third, the paper focuses on the call blocking, channel selection, and assignment. However, in most of the previous works on spectrum, allocation and sharing[12]–[17] allocations are done in unlicensed band in a distributed manner. In this work, spectrum sharing covers with opportunistic spectrum sharing in inter and intra network spectrum sharing in a heterogeneous wireless networks.

## 3. Cognitive Radio Process

CR is a radio that is able to sense the spectral environment over a wide frequency band and exploit this information to opportunistically provide wireless links that best meet the user communications requirements. CR provides the real-time interaction with its environment.

This provides a way to dynamically adapt to the dynamic radio environment and the radio analyzes the spectrum characteristics and changes the parameters among the users that share the available spectrum.

The goals of adaptation include spectral efficiency, minimizing interference to other CRs, coexistence of licensed users, etc. The environmental parameters that are continually sensed for adaptation include occupied radio frequency bands, user traffic, network state, etc. The CR consists of three major components:

1) RF sensing. It refers to the estimation of the total interference in the radio environment, detection of the spectrum holes (or unused bands), estimation of the channel state information (i.e. SINR), and prediction of channel capacity for use by the transmitter[18].

2) Cognition/management. It refers to the spectrum management which controls the opportunistic spectrum access, traffic shaping, QoS provisioning, etc.

3) Control action. It refers to the transmit-power control, adaptive modulation and coding, and transmission rate control.

The main features of CR are listed as below[19], [20].

### 3.1 Spectrum Sensing

It detects the unused spectrum and shares it without harmful interference with other users. It is an important requirement of the CR network to sense the spectrum holes. Primary users detection is found to be the most efficient way to detect the spectrum holes. Some of the spectrum sensing techniques can be classified as follows[21].

*A. Transmitter Detection*

In this category, CR must have the capability to determine if a signal from a primary transmitter is locally present in a certain spectrum. Some proposed approaches in this category are: matched filtering detection, energy detection, waveform based sensing, cyclostationary based sensing, etc.

*B. Cooperative Detection*

In this category, it decreases the probabilities of misdetection and false alarm considerably. It can also solve the hidden primary user problem and decrease the sensing time.

*C. External Detection*

In this category, an external agent performs the sensing and broadcasts the channel occupancy information to CR. The main advantage is to overcome the hidden primary user problem as well as the uncertainty due to shadowing and fading. As the CR does not spend time for sensing, spectrum efficiency is increased.

### 3.2 Spectrum Management

It is the task of capturing the best available spectrum to meet users' requirements. CR should decide on the best spectrum band to meet the QoS requirements over all available spectrum bands, therefore spectrum management functions are required for CRs. This category can be classified as follows.

*A. Spectrum Analysis*

In this technique, each spectrum hole should be characterized considering not only the time-varying radio environment and but also the primary user activity.

*B. Spectrum Decision*

When all the analysis of spectrum band is done, appropriate spectrum band should be selected for the current transmission considering the QoS requirements and the spectrum characteristics. According to users' requirement on the data rate, bandwidth is determined, and according to the decision rule, an appropriate spectrum band is chosen.

### 3.3 Spectrum Mobility

It is a process when the CR user exchanges its frequency of operation. CR networks target to use the spectrum in a dynamic manner by allowing the radio terminals to operate in the best available frequency band, maintaining seamless communication requirements during the transition to better spectrum.

### 3.4 Spectrum Sharing

It refers to providing the fair spectrum scheduling method, one of the major challenges in the open spectrum usage is the spectrum sharing. CRs have the capability to sense the surrounding environments and allow intended secondary user to increase QoS by opportunistically using the unutilized spectrum holes. If a secondary user senses the available spectrum, it can use this spectrum after the primary licensed user vacates it.

In our work, the above mentioned features of CR are applied in order to make spectrum sharing among service providers with the help of CR.

## 4. Dynamic Spectrum Access

Dynamic spectrum access techniques allow the cognitive radio to operate in the best available channel. Radio spectrum is considered as a scarce resource with the growing demand for spectrum-based services because a major portion of the spectrum has been allocated for licensed wireless applications.

The first step in dynamic spectrum access is the detection of unused spectral bands. Therefore, CR device is used for measuring the RF energy in a channel to determine whether the channel is idle or not. But, this approach has a problem in that wireless devices can only sense the presence of a primary user (PU) if and only if the energy detected is above a certain threshold. Taxonomy of dynamic spectrum access has been depicted in the Fig. 1.

Generally, dynamic spectrum access can be categorized into three models, namely[22]–[25]:

1) Dynamic exclusive use model;
2) Open sharing model;
3) Hierarchical access model.

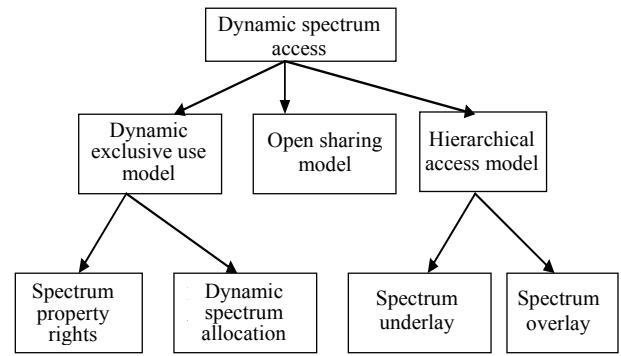

Fig. 1. Taxonomy of dynamic spectrum access.

### 4.1 Dynamic Exclusive Use Model

This model manages spectrum in the finer scale of time, space, frequency and use dimensions so at any given point in space and time, only one operator has exclusive right to the spectrum but the identity of the owner and type of use can change.

This model maintains the basic structure of the current spectrum regulation policy: spectrum bands are licensed to services for exclusive use. The main idea is to introduce flexibility to improve spectrum efficiency. Two approaches have been proposed under this model: spectrum property rights and dynamic spectrum allocation. The former approach allows licensees to sell and trade spectrum and freely choose technology.

The other approach, dynamic spectrum allocation, aims to improve spectrum efficiency through dynamic spectrum assignment by exploiting the spatial and temporal traffic statistics of different services. Similar to the current static spectrum allotment policy, such strategies allocate, at a given time and region, a portion of the spectrum to a radio access network for its exclusive use.

Based on an exclusive-use model, these approaches cannot eliminate white space in the spectrum resulting from the bursty nature of wireless traffic.

### 4.2 Open Sharing Model

This model is called as spectrum commons. This model employs open sharing among peer users as the basis for managing a spectral region. Centralized and distributed spectrum sharing strategies have been initially investigated to address technological challenges under this model.

### 4.3 Hierarchical Access Model

This model is considered to be the same as the exclusive-use model. The basic idea is to open licensed spectrum to secondary users and limit the interference perceived by primary users (licensees). This model is first to describe simultaneous shared use of spectrum wherein there is a primary licensed owner of the spectrum band and multiple secondary users opportunistically share the band.

Spectrum sharing between primary and secondary users utilizes spectrum underlay and spectrum overlay approaches.

Sometimes the hierarchical access model can be categorized under the open sharing model because spectrum sharing between primary and secondary users is fundamentally different from spectrum sharing among peer users in both technical and regulatory aspects. But in this paper, both the papers got separated.

*A. Spectrum Underlay*

The underlay approach allows primary and secondary user transmission simultaneously in the manner of ultra wideband (UWB) systems. To protect the primary users, system provides a spectral mask to secondary signals so that the interference generated by the secondary devices is below the acceptable noise floor for the primary users of the spectrum. This technique allows communication over short range.

*B. Spectrum Overlay*

With the help of this technique, primary and secondary users are allowed to transmit data simultaneously. The alternate for overlay system is that for secondary communication secondary users use their power and the remaining power to relay primary transmission. Spectrum overlay approach targets at spatial and temporal unused radio spectrum called white space by allowing secondary users to identify and exploit local and international instantaneous spectrum availability in non-intrusive manner. Therefore, exclusive knowledge about other signals in the spectrum is necessary.

In this work, we applied the spectrum overlay concept in order to identify the spectrum availability for secondary users.

# 5. Proposed Spectrum Sharing Techniques

## 5.1 Process of Spectrum Sharing

During the peak hours, the communication of the users will be blocked due to the number of active users is greater than the maximal number of users, i.e., the infrastructure is found to be over-loaded (because of the channel scarcity). At the same time, the infrastructure of the other service providers might be in the under-loaded status. Hence, the available channels of the under-loaded service providers can be utilized by the overloaded service providers. This may vary according to the services offered by the customers. So this may lead to the difference in the traffic of the active users across service providers. Here, both of these two service providers operate cell-based wireless networks. Therefore, implementing spectrum sharing among service providers would highly improve the spectrum efficiency and it also reduces the call blocking rate and co-channel interference[26].

In order to reduce the co-channel interference and to remove the need of equipping CRs in each infrastructure of service providers, we propose a method that implements a spectrum sharing among service providers via CR nodes in a long-term spectrum assignment scheme. These CR nodes are distributed regularly within an area of interest. Each CR node senses the surrounding environment and monitors the channel usage within its sensing range of different service providers. To avoid co-channel interference, CR node provides the list of the channel availability of each cell for the overloaded infrastructure of service provider. CR nodes are connected to each other via wire or they communicate wirelessly to form a network. This network is called as spectrum management network and it coexists with the wireless networks operated by different service providers.

In this technique, only a limited number of CR nodes are required, and neither users nor service providers need to sense the environment for available spectrum. In the proposed technique, each user subscribes to a specific service provider that is assigned fixed frequency bands. When one or more infrastructures (e.g. base stations) of a service provider are overloaded, they use extra available channels (for communication) which are licensed to other service providers. The overloaded infrastructures obtain the channel availability information from surrounding CR nodes. CR nodes are deployed to estimate the channel utilization and provide the channel usage information for infrastructures upon their requests. The infrastructures process the information received from CR nodes to select the optimum channels based on the channel associated metrics such as interference level, cost, and the probability of channel being available for certain time duration.

## 5.2 Operations of CR Nodes

Suppose, if the infrastructure of one service provider is found to be overloaded, it sends the request to the adjacent CR nodes regarding the channel usage availability in its coverage area[26]. It gives the list of available channel from the relevant adjacent CR nodes and it also provides the relevant information associated with each channel such as the average signal to noise-plus interference[27]. The overloaded infrastructure now selects the proper channel to use. This is clearly identified in Fig. 2.

Thus, users can communicate with the overloaded infrastructure over the new channels after they are informed with the channel availability. This requires both infrastructure and users to be equipped with radios capable of operating over different frequency bands. If the infrastructure is not found to be overloaded or the traffic is free, the channels should be released, i.e. users would stop using these channels.

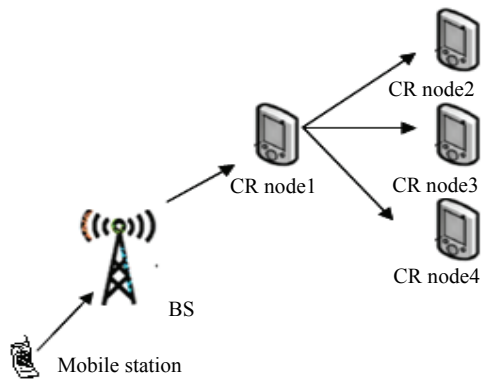

Fig. 2. CR nodes sensing the available channels.

The main operation of CR nodes is to periodically sense the environment and estimate the channel utilization within their sensing range. Once it receives the channel availability information from infrastructures, CR nodes would response to these infrastructures with a set of available channels and relevant information. The main challenges we are facing during the spectrum sharing are:

(1) How to decide the availability of a channel – for a CR node ?

(2) How to select the optimum channels for usage based on the information provided by CR nodes—for the infrastructure?

The solutions are given by the determination of channel availability in CR nodes and in an infrastructure, i.e. CR nodes senses the unused channels and an overloaded infrastructure would receive the channel usage information from its adjacent CR node after request.

Thus, implementing those CR nodes and deploying them as a network should be done with the negotiation between service providers based on the cost and the network management policy. This successful deployment provides the risk and the cost of operating the CR network. It is found to be a challenging task of deploying the CR network with the coexistence with the current wireless networks in a geographical situation. It also increases the cost and complexity of wireless network management. If CR nodes communicate with each other wirelessly, then it requires extra wireless resources and it also increases the overhead of wireless networks.

### 5.3 Simulation Results of Spectrum Sharing

If the infrastructure of one service provider is found to be overloaded, it sends the requests to the adjacent CR nodes regarding the availability of channel. Channel availability can be determined by sending service request to the BS. BS receives service request from the mobile nodes and it will send the channel request to the CR node. Fig. 3 shows the BS before sending request to the CR nodes.

CR node receives the channel request and sends the broadcast message to the adjacent CR node. A neighbor CR node receives the broadcast message and also sends the available channel list to the BS. CR node and its neighbors update the channel availability list and send response to the BS. Fig. 4 shows the sending and receiving the request and response from CR nodes.

If BS receives the response from the CR node, it selects the available channel and sends service reply with the allocated channel to the mobile nodes. This shows the maximum utilization of a channel and it also offers several services such as internet service, call service, multimedia service, and so on to the mobile nodes. Fig. 5 shows the maximum utilization of channel by sending the response to the BS regarding the channel availability.

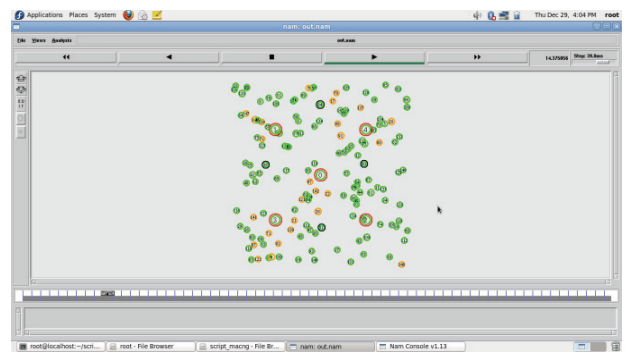

Fig. 3. Before sending request to the CR nodes.

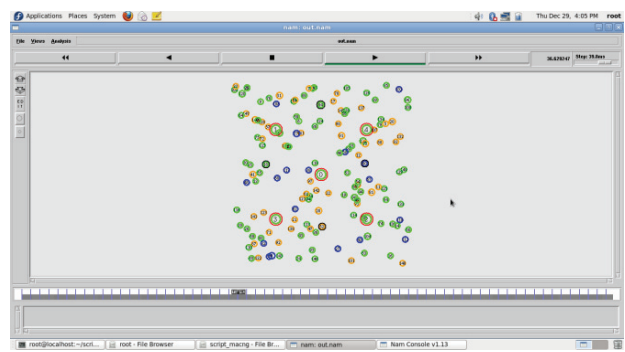

Fig. 4. Sending request to the CR nodes and receiving response from the CR nodes.

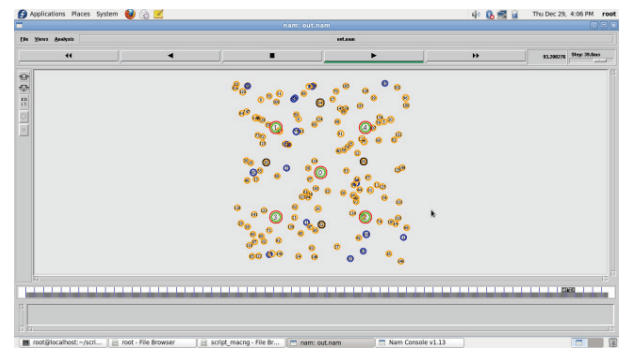

Fig. 5. Response from CR nodes.

Our scheme identifies 'available' channel list for each CR node. Such a list shows which channel is available to use depending on the distance between the CR node and high frequency band. Within a given neighborhood, multiple CR nodes may contend for access to one or more of the available channels.

## 6. Performance Metrics

In this section, we discuss about the performance metrics to study the impact of spectrum sharing on the service providers including call blocking rate, system efficiency, revenue efficiency, etc.

### 6.1 Blocking Rate

The call blocking rate $R_{BL}$ is defined as the ratio of total blocked calls over total calls processed by all service providers and corresponds to

$$R_{BL} = \lim \frac{n_{BL}^{(total)}(t)}{n_{processed}^{(total)}(t)}$$

where total blocked calls at time $t$ by all service providers is given by

$$n_{BL}^{(total)}(t) = \sum_{i=1}^{n_{sp}} n_{BL}^{(i)}(t)$$

and the total calls processed is

$$n_{processed}^{(total)}(t) = \sum_{i=1}^{n_{sp}} n_{processed}^{(i)}(t)$$

where $n_{sp}$ is the number of service providers. Here, the call would be blocked, if all the service providers are overloaded.

### 6.2 System Efficiency

The system efficiency $\eta_{sys}^{(i)}$ is defined as probability efficiency metric for service provider is determined by the processed traffic intensity and the total traffic loaded to service provider within the observation time. Thus, $\eta_{sys}^{(i)}$ is calculated by

$$\eta_{sys}^{(i)} = \frac{E_p^{(i)}}{E_{in}^{(i)}}$$

where $E_p^{(i)}$ is the processed traffic intensity in Erlang for service provider $i$ and $E_{in}^{(i)}$ is the total traffic loaded to the service provider $i$ within the observation time $t$.

### 6.3 Spectrum Utilization Efficiency

The spectrum efficiency $\eta_s^{n_{(sp)}}$ is defined as the ratio of average busy channels over total channels owned by service providers. It corresponds to

$$\eta_s^{n_{(sp)}} = \lim \frac{1}{t} \int_0^t \frac{n_{busy}^{n_{(sp)}}(t)}{N_{ch-total}^{n_{(sp)}}(t)} dt$$

where $n_{busy}^{n_{(sp)}}(t)$ is the number of channels used at time $t$ for service provider $n_{(sp)}$ and $N_{ch-total}^{n_{(sp)}}(t)$ is the total number of total channels owned by service provider $n_{(sp)}$. Higher spectrum efficiency is estimated because the call blocking rate is lower; thus more calls can contribute to the spectrum utilization.

### 6.4 Cost (Revenue) Efficiency

Within the observation time, cost is determined by the number of processed calls and the length of call holding time. We define the metric $c_e^{(i)}$ to reflect the cost efficiency. $c_e^{(i)}$ is the ratio of the cost earned within the observation time $t$ over total input traffic intensity for service provider sp, and it is defined as

$$c_e^{(i)} = \frac{c^{(i)}}{E^{(i)}} = \frac{\alpha^{(i)} E_p t^{(i)}}{E^{(i)}} = \alpha^{(i)} t^{(i)} \eta_s^{(sp)}$$

where $\alpha^{(i)}$ is the unit price ($/second/channel) for service provider sp and $c^{(i)}$ is the average income within the observation time.

## 7. Simulation Results

In this section, we present simulation results on the performance of our proposed sensing framework. Channel assignment mechanisms in the traditional multi-channel wireless networks typically select the 'best' channel for a given transmission. In the proposed work, we are choosing the available channel with the high probability and high-frequency band. To generate utility performance measures, we assume:

1) Maximal five service providers share their spectra, and 150 nodes are chosen. Maximum limit of user per channel is 10.
2) Call arrival of each service provider is the heterogeneous process.
3) Traffic rates are correlated jointly-Gaussian random variables.
4) The infrastructures for different service providers are located at the same position and the cell radii is also the same.
5) The CR nodes are present at the vertices of the cells of the service providers.
6) Each CR node has the ability of sensing its range within the coverage limits.
7) CR nodes have the capability of detecting all the

available channels that are licensed to the other service providers.

8) Channel parameters such as the interference level, the probability of being available for a given time period and cost are the same for all available channels.

We conduct simulations to verify the potential of the call arrival rate for different service providers in terms of utility performance measures.

Table 1 shows the values for the number of active users for different service providers.

As the number of active users is reduced, the traffic rate becomes higher as illustrated in the Fig. 6. Thus, when the traffic is increased, call blocking is reduced which indicates high correlation between the call blocking and service provider. If the call blocking is reduced among the correlated service providers, then there would be an increase in the active users.

Table 2 shows the values for calculating the call block for different service providers.

Table 1: Values for active users

| S.No. | No. of nodes | Active users for service providers | | |
|---|---|---|---|---|
| | | 1_SP_ _ | 3_SP_ _ | 5_SP_ _ |
| 1 | 20 | 5 | 1 | 0 |
| 2 | 40 | 6 | 1 | 0 |
| 3 | 60 | 11 | 3 | 1 |
| 4 | 80 | 19 | 5 | 2 |
| 5 | 100 | 21 | 6 | 3 |

As illustrated in Fig. 7, the call blocking rate is highly correlated with different service providers. Thus, the traffic rate increases along with the call blocking rate.

Table 3 predicts the traffic load between the active users among different service providers.

Fig. 8 shows that as the call blocking is reduced then the traffic load for five service providers is the minimum.

Interference is the key factor that limits the performance of wireless networks. Spectrum managers are concerned with managing interference and in establishing the methods, techniques, information and processes need to protect the users from interference. So, interference arises in radio systems when a transmitter's ability to communicate with its intended receiver(s) is limited because of the transmissions of other transmitters. The problem may be thought of as arising from the limitations of the receiver: better receivers are able to extract the desired signal from a noisy environment of background radiation and other transmitters.

$$\text{interference} = |\text{frmax} - \text{frmin}|.$$

Table 4 shows the communication range values for interference among different service providers.

As the call arrival rate increases then there will be a minute decrease in the interference. This is shown in Fig. 9.

Table 5 shows the values for the channel utilization among different service providers.

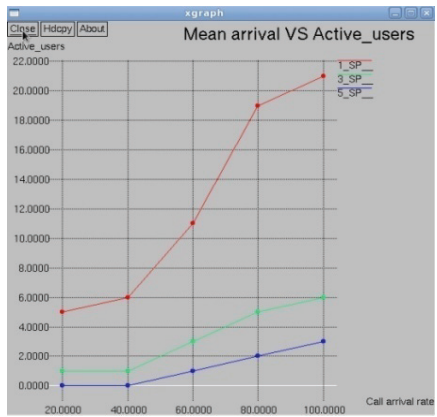

Fig. 6. Mean arrival vs. active users.

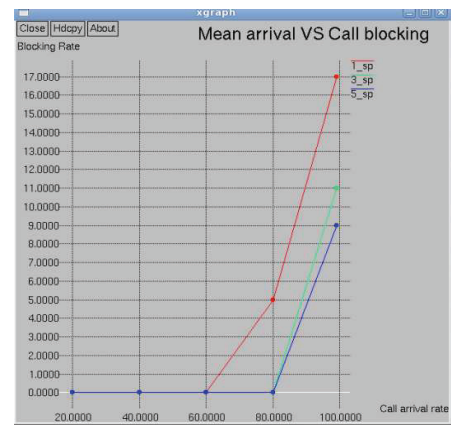

Fig. 7. Mean arrival vs. call blocking.

Table 2: Values for call blocking

| S.No. | No. of nodes | Number of call blocking rate | | |
|---|---|---|---|---|
| | | 1_sp | 3_sp | 5_sp |
| 1 | 20 | 0 | 0 | 0 |
| 2 | 40 | 0 | 0 | 0 |
| 3 | 60 | 0 | 0 | 0 |
| 4 | 80 | 5 | 0 | 0 |
| 5 | 100 | 17 | 11 | 9 |

Table 3: Values for traffic load

| S.No. | No. of nodes | Values for traffic load | | |
|---|---|---|---|---|
| | | 1_sp_ _ | 3_sp_ _ | 5_sp_ _ |
| 1 | 20 | 16.6667 | 13.3333 | 6.6667 |
| 2 | 40 | 20 | 16.6667 | 13.3333 |
| 3 | 60 | 36.6667 | 30 | 20 |
| 4 | 80 | 63.3333 | 50 | 40 |
| 5 | 100 | 70 | 63.3333 | 50 |

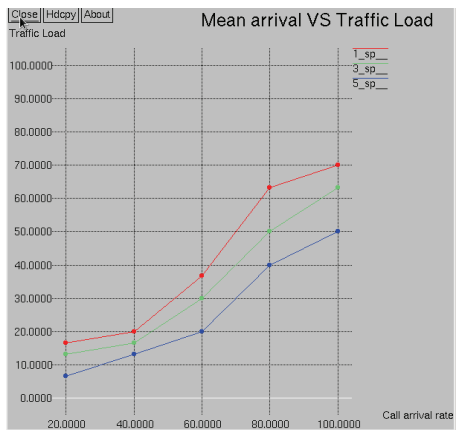

Fig. 8. Mean arrival vs. traffic load.

Table 4: Values for interference

| S.No. | No. of nodes | Communication range values | | |
|---|---|---|---|---|
| | | 1_sp_ | 3_sp_ | 5_sp_ |
| 1 | 20 | 1188.02 | 1185.52 | 1180.52 |
| 2 | 40 | 1170.52 | 1168.02 | 1165.52 |
| 3 | 60 | 1163.02 | 1158.02 | 1150.52 |
| 4 | 80 | 1163.02 | 1153.02 | 1145.52 |
| 5 | 100 | 1148.02 | 1143.02 | 1133.02 |

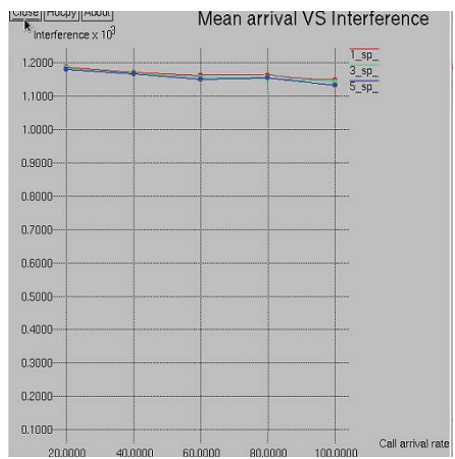

Fig. 9. Mean call arrival vs. interference.

Table 5: Values for channel utilization

| S.No. | No. of nodes | Signals for service providers | | |
|---|---|---|---|---|
| | | 1_Sp | 3_Sp | 5_Sp |
| 1 | 20 | 0.62353 | 0.202697 | 0.12162 |
| 2 | 40 | 0.60195 | 0.20207 | 0.12124 |
| 3 | 60 | 0.6068 | 0.208041 | 0.12483 |
| 4 | 80 | 0.61792 | 0.199994 | 0.12192 |
| 5 | 100 | 0.61792 | 0.201009 | 0.12599 |

As the mean call arrival increases and call blocking decreases, the channel utilization also increases. It is shown in the Fig. 10. Higher spectrum efficiency is estimated because the call blocking rate is lower; thus, more calls can contribute to the spectrum utilization.

Table 6 shows the values of channel capacity to know the efficiency of the spectrum among different service providers.

Fig. 11 shows that, at high-traffic rates, the system efficiency is lower when the traffic rates of different service providers are highly correlated. When the correlation is lower, based on Fig. 11, as the dropped calls decrease, thus, the total processed calls increase. The system efficiency decreases when the traffic rate is beyond the system capacity.

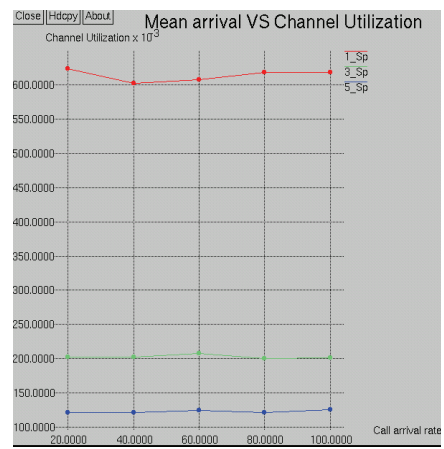

Fig. 10. Mean arrival vs. spectrum efficiency.

Table 6: Values for system efficiency

| S.No. | No. of nodes | System efficiency | | |
|---|---|---|---|---|
| | | 1_SP | 3_SP | 5_SP |
| 1 | 20 | 1 | 1 | 1 |
| 2 | 40 | 1 | 1 | 1 |
| 3 | 60 | 1 | 1 | 1 |
| 4 | 80 | 0.97561 | 1 | 1 |
| 5 | 100 | 0.92166 | 0.947867 | 0.95694 |

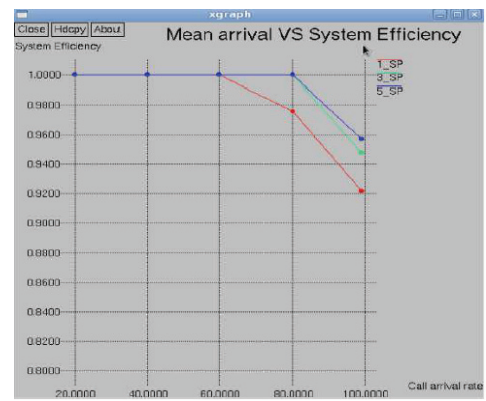

Fig. 11. Mean arrival vs. system efficiency.

# 8. Conclusions

The spectrum assigned to different service providers is not properly utilized with the same frequency. So some service providers may try to use the allocated spectrum fully, and even they need more spectrums, which may not be used by service providers fully. This paper provides an offer (way) to utilize and also to share the licensed spectrum among the service providers if they are under utilized.

Here, we discussed about the operations of CR nodes and infrastructures of service providers for spectrum sharing. From the proposed work, we can sense the range of CR node and we decide and select the optimal channel for spectrum sharing. We have derived general formulae for the interference, call blocking rate, spectral utilization, and cost.

In addition to this work, we also define spectral efficiency as performance metric for spectrum sharing among service provides in order to reach an efficient spectrum utilization. This technique removes the need of sensing spectrum in each user. Thus, it reduces the cost, complexity, and battery power consumption of user devices.

Furthermore, the performance of the wireless network with opportunistic spectrum sharing techniques is analyzed. Spectral utilization and probability efficiency of sensing are increased, which minimizes the interference and reduces the call blockage.

In this work, interference is found with minute difference. If we apply OFDM, we can avoid interference among the service providers. Avoiding interference using OFDM will be applied in our future work.

# References

[1] R. Kaniezhil, C. Chandrasekar, and S. N. Rekha, "Channel selection for spectrum sharing using CR nodes," *Int. Proc. of Computer Science and Information Technology*, vol. 20, pp. 93–98, Dec. 2011.

[2] R. Kaniezhil, C. Chandrasekar, and S. N. Rekha, "Performance evaluation of QoS parameters in spectrum sharing using SBAC algorithm," in *Proc. of IEEE Int. Conf. on Advances in Engineering, Science and Management*, Nagapattinam, 2012, pp. 755–760.

[3] R. Kaniezhil and C. Chandrasekar, "Spectrum sharing in a long term spectrum strategy via cognitive radio for heterogeneous wireless networks," *Int. Journal on Computer Science and Engineering*, vol. 4, no. 6, pp. 982–995, 2012.

[4] R. Kaniezhil, Dr. C. Chandrasekar, "Multiple service providers sharing spectrum using cognitive radio in wireless communication networks," *Int. Journal of Scientific & Engineering Research*, vol. 3, no. 3, pp. 1–6, 2012.

[5] J. Acharya and R. D. Yates, "A framework for dynamic spectrum sharing between cognitive radios," in *Proc. of IEEE ICC 2007*, Glasgow, 2007, pp. 5166–5172.

[6] W. Wang, T.-J. Lv, Z.-Y. Ren, L. Gao, and W.-D. Liu, "A novel spectrum sharing algorithm based on the throughput in cognitiveradio networks," in *Proc. of 2010 IEEE 72nd Vehicular Technology Conf.*, Ottawa, 2010, pp. 1–5.

[7] M. Alicherry, R. Bhatia, and L. Li, "Joint channel assignment and routing for throughput optimization in multi-radio wireless mesh networks," in *Proc. of MobiCom'05*, Cologne, 2005, pp. 1–15.

[8] F. F. Digham, "Joint power and channel allocation for cognitive radios," in *Proc. of IEEE WCNC 2008*, Las Vegas, 2008, pp. 882–887.

[9] J. Tang, S. Misra, and G.-L. Xue, "Joint spectrum allocation and scheduling for fair spectrum sharing in cognitive radio wireless networks," *Computer Networks Journal*, vol. 52, no. 11, pp. 2148–2158, 2008.

[10] Q.-W. Wang and H.-T. Zheng, "Route and spectrum selection in dynamic spectrum networks," in *Proc. of the 3rd IEEE Consumer Communications and Networking Conf.*, Las Vegas, 2006, pp. 1–5.

[11] I. F. Akyildiz, W.-Y. Lee, M. C. Vuran, and S. Mohanty, "Next generation/dynamic spectrum access/cognitive radio wireless networks: a survey," *Computer Networks*, vol. 50, pp. 2127–2159, Sep. 2006.

[12] R. Etkin, A. Parekh, and D. Tse, "Spectrum sharing for unlicensed bands," in *Proc. of IEEE Int. Symposium on New Frontiers in Dynamic Spectrum Access*, Baltimore, 2005, pp. 251–258.

[13] M. Buddhikot, P. Kolody, S. Miller, K. Ryan, and J. Evans, "DIMSUMNet: new directions in wireless networking using coordinated dynamic spectrum access," in *Proc. of IEEE WoWMoM 2005*, Taormina, 2005, pp. 7885.

[14] W.-Y. Lee and I. F. Akyildiz, "Joint spectrum and power allocation for inter-cell spectrum sharing in cognitive radio networks," in *Proc. of IEEE DySPAN*, Chicago, 2008, pp. 1–12.

[15] W.-Y. Lee and I. F. Akyildiz, "A spectrum decision framework for cognitive radio networks," *IEEE Trans. on Mobile Computing*, vol. 10, no. 2, pp. 161–174, 2011.

[16] J. Huang, R. A. Berry, and M. L. Honig, "Spectrum sharing with distributed interference compensation," in *Proc. IEEE DySPAN 2005*, Baltimore, 2005, pp. 88–93.

[17] L. Zhang, Y. Liang, and Y. Xin, "Joint beamforming and power allocation for multiple access channels in cognitive radio networks," *IEEE Journal on Selected Areas in Communications*, vol. 26, pp. 38–51, Jan. 2008.

[18] J. Jia, Q. Zhang, and X. Shen, "HC-MAC: a hardware constrained cognitive MAC for effiicient spectrum management," *IEEE Journal on Selected Areas in Communications*, vol. 26, pp. 106–117, Jan. 2008.

[19] K.-C. Chen and R. Prasad, *Spectrum Management of Cognitive Radio Networks*, New York: John Wiley & Sons Ltd, 2009, pp. 335–355.

[20] Z. Tabakovic, S. Grgic, and M. Grgic, "Dynamic spectrum access in cognitive radio," in *Proc. of the 51st Int. Symposium ELMAR-2009*, Croatia, 2009, pp. 245–248.

[21] T. Y¨ucek and A. H¨useyin, "A survey of spectrum sensing algorithms for cognitive radio applications," *IEEE*

*Communications Surveys & Tutorials*, vol. 11, no. 1, 2009, pp. 116–130.
[22] Q. Zhao and A. Swami, "A survey of dynamic spectrum access: signal processing and networking perspectives," in *Proc. of IEEE ICASSP 2007*, Honolulu, 2007, pp. 1349–1352.
[23] M. M. Buddhikot, "Understanding dynamic spectrum access: models, taxonomy and challenges," in *Proc. of IEEE DySPAN 2007*, Dublin, 2007, pp. 649–663.
[24] T. J. Harrold, L. F. Wang, M. A. Beach, G. Salami, A. Yarmohammad, and O. Holland, "Spectrum sharing and cognitive radio opportunities for efficiency enhancement," in *Proc. of Int. Conf. on Ultra Modern Telecommunications & Workshops*, St. Petersburg, Russia, pp. 1–8.
[25] I. F. Akyildiz, W.-Y. Lee, and K. R. Chowdhury, "CRAHNs: cognitive radio ad hoc networks," *Elsevier Ad Hoc Networks*, vol. 7, pp. 810–836, 2009, doi: 10.1013/j.adhoc.2009.01.001.
[26] V. Brik, E. Rozner, S. Banarjee, and P. Bahl, "DSAP: a protocol for coordinated spectrum access" in *Proc. of IEEE DySPAN 2005*, Baltimore, 2005, pp. 611–614.
[27] S. Haykin, "Cognitive radio: brain-empowered wireless communications," *IEEE Journal on Selected Areas in Communications*, vol. 23, no. 2, pp. 201–220, 2005.

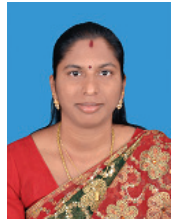
**R. Kaniezhil** received the B.Sc. degree from the University of Madras in 1998. She received her MCA and M.Phil degrees from Periyar University and Annamalai University, in 2001 and 2007, respectively. She is currently pursuing Ph.D. degree with the Department of Computer Science, Periyar University, Salem, India. Her research interests include mobile computing, spectral estimation, and wireless networking.

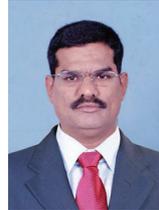
**C. Chandrasekar** received his Ph.D. degree from Periyar University. He is working as an associate professor with the Department of Computer Science, Periyar University, Salem. His areas of interest include wireless networking, mobile computing, computer communications and networks. He is a research guide with various universities in India. He has published more than 100 technical papers on various national & international conferences and 63 journals.